\documentclass[fleqn,usenatbib]{mnras}
\usepackage{newtxtext,newtxmath}
\usepackage[T1]{fontenc}
\usepackage{ae,aecompl}
\usepackage{graphicx}	
\usepackage{amsmath}	
\usepackage{threeparttable}
\usepackage{multirow}
\usepackage{romannum}

\usepackage{appendix}

\title[Application of the emission model to PAHs and C$_{60}$]{Polycyclic Aromatic Hydrocarbon emission model in photodissociation regions\\
\begin{LARGE} 
  \Romannum{2}: Application to the Polycyclic Aromatic Hydrocarbon and fullerene emission in NGC 7023
\end{LARGE} }

\author[A. Sidhu et al.]{
Ameek Sidhu,$^{1,2}$\thanks{E-mail: asidhu92@uwo.ca}
A.G.G.M. Tielens,$^{3,4}$
Els Peeters$^{1,2,5}$
and Jan Cami$^{1,2,5}$
\\
$^{1}$Department of Physics \& Astronomy, University of Western Ontario, London, ON, N6A 3K7, Canada\\
$^{2}$Institute for Earth and Space Exploration, University of Western Ontario, London, ON, N6A 3K7, Canada\\
$^{3}$Leiden Observatory, Leiden University, Niels Bohrweg 2, 2333 CA Leiden, Netherlands \\
$^{4}$ Department of Astronomy, University of Maryland, College Park, MD 20742, USA \\
$^{5}$ SETI Institute, 339 Bernardo Avenue, Suite 200, Mountain View, CA 94043, USA
}

\begin{document}
\label{firstpage}
\pagerange{\pageref{firstpage}--\pageref{lastpage}}
\maketitle

\begin{abstract}
We present a charge distribution-based emission model that calculates the infrared spectrum of fullerenes (C$_{60}$). Analysis of the modelled spectrum of C$_{60}$ in various charge states shows that the relative intensity of the features in the 5-10 $\mu$m versus 15-20 $\mu$m can be used to probe the C$_{60}$ charge state in interstellar spectra. We further used our model to simulate emission from polycyclic aromatic hydrocarbons (PAHs) and C$_{60}$ at five positions in the cavity of reflection nebula NGC~7023. Specifically, we modelled the 6.2/11.2 band ratio for circumcoronene and circumcircumcoronene and the 7.0/19.0 band ratio for C$_{60}$ as a function of the ionization parameter $\gamma$. A comparison of the model results with the observed band ratios shows that the $\gamma$ values in the cavity do not vary significantly, suggesting that the emission in the cavity does not originate from locations at the projected distances. Furthermore, we find that the C$_{60}$ derived $\gamma$ values are lower than the PAH-derived values by an order of magnitude. We discuss likely scenarios for this discrepancy. In one scenario, we attribute the differences in the derived $\gamma$ values to the uncertainties in the electron recombination rates of PAHs and C$_{60}$. In the other scenario, we suggest that PAHs and C$_{60}$ are not co-spatial resulting in different $\gamma$ values from their respective models. We highlight that experiments to determine necessary rates will be required in validating either one of the scenarios.            
\end{abstract}


\section{Introduction}
Polycyclic Aromatic Hydrocarbons (PAHs) and fullerenes are two families of large aromatic molecules that are widespread and abundant in the Universe \citep{Sellgren:1983, Leger:1984, Allamandola:1985, Galliano:2008, Cami:2010, Sellgren:2010}. While PAHs are planar molecules composed of fused benzene rings decorated with hydrogen atoms at their edges, fullerenes are closed cage structures of carbon. Buckminsterfullerene, C$_{60}$, is the most famous fullerene molecule. It has been unequivocally detected in the Universe and is the largest molecule discovered to date \citep[e.g.][]{Cami:2010, Sellgren:2010, Berne:2013, Campbell:2015, Cordiner:2019}.

Despite their distinct molecular structures, PAHs and C$_{60}$ share a common excitation and emission mechanism \citep{Tielens:2021}. PAHs and C$_{60}$ are both strong ultraviolet (UV) photon absorbers that release the absorbed energy via their vibrational modes in the mid-infrared (MIR). Emission from these molecules typically originates from photodissociation regions (PDRs), the transition regions between the ionized gas and the molecular clouds where far-ultraviolet (FUV; 6-13.6 eV) photons drive the physics and chemistry of the gas. \citet{Berne:2012} have hypothesized that these two molecular families are possibly linked by the mechanism of C$_{60}$ formation from the UV processing of PAHs. However, several questions remain about the formation and destruction of PAHs and C$_{60}$ and how they are related to each other.

In \citet{Sidhu:2022}, hereafter referred to as Paper I, we presented an emission model that calculates the charge distribution of PAHs using experimentally measured and theoretically calculated molecular properties of PAHs and combines that charge distribution with the emission model of PAHs to calculate the total PAH emission in astrophysical environments. In Paper I, we demonstrated that the charge distribution-based emission model could explain PAH observations in five different PDRs. In this paper, we extend our model to C$_{60}$ given the similarity in the excitation mechanism of PAHs and C$_{60}$. We model the emission of PAHs and C$_{60}$ at five different locations in the cavity of the reflection nebula NGC~7023 and compare our results with observations in order to provide more insight into the relation between C$_{60}$ and PAHs.

This paper is organized as follows. In section~\ref{sec:NGC_7023}, we describe the astrophysical environment NGC~7023. In section~\ref{sec:Method}, we apply our emission model to PAHs and C$_{60}$, and in section~\ref{sec:comparison}, we compare the model results to the observations. In section~\ref{sec:Discussion}, we discuss the disagreement between model results for C$_{60}$ and PAH observations. Finally, in section~\ref{sec:Summary}, we provide a summary of this work.    

\section{NGC 7023}
\label{sec:NGC_7023}
\begin{figure}
    \centering
    \includegraphics[scale = 0.35]{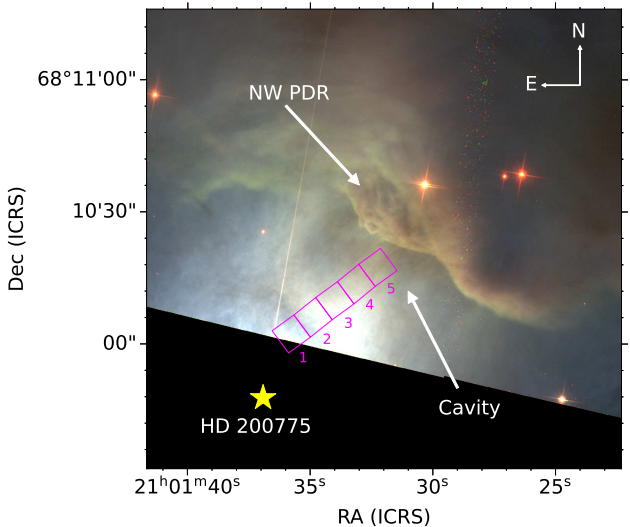}
    \caption{The three-colour Hubble Space Telescope ACS image of NGC 7023. The red channel corresponds to the combined optical H$\alpha$ (658 nm) and infrared I-band (850 nm) filters, the green channel to the optical V-band (625 nm) filter, and the blue channel to the optical B-band (475 nm) filter. The yellow star indicates the location of the illuminating star HD~200775. The figure clearly shows the cavity carved out by the star, the walls of which are outlined in the image tracing PDR emission. The cavity and one of the PDRs, the NW PDR, are annotated in the figure. The positions of the cavity studied in this paper have been labelled from 1 to 5 and are shown in pink rectangles.}
    \label{fig:spatial_map_NGC_7023}
\end{figure}

NGC~7023 is a bright well-studied reflection nebula located at a distance of $\sim$361 pc \citep{gaia16, gaia18b}. It is illuminated by a spectroscopic binary Herbig B3Ve - B5 star HD~200775 having an effective temperature (T$_{\rm eff}$) of 17000 K \citep{VandenAncker:1997, Witt:2006, Alecian:2013}. UV, optical, and IR observations have shown that the illuminating star blew away the surrounding gas, leaving an hourglass shaped cavity almost completely devoid of gas \citep{Rogers:95, Fuente:1996, Gerin:1998}. The edges of the cavity are delineated with PDRs $\sim$42$''$ North-West (NW), 55$''$ South-West (SW), and 150$''$ East (E) of the illuminating star. Fig.~\ref{fig:spatial_map_NGC_7023} shows the NW PDR and the cavity blown by the illuminating star as observed in the three-colour Hubble Space telescope ACS image of NGC~7023. It is worth pointing out that the cavity shows a faint H$\alpha$ glow, which rather than revealing the presence of ionized gas, is due to reflection by the dust of the H$\alpha$ emission associated with accretion onto the central star \citep{Witt:2006}.

NGC~7023 is one of the first objects in which both the mid-IR PAH and C$_{60}$ emission was observed \citep[][]{Sellgren:1983, Sellgren:2010, Berne:2013}. While PAHs have been associated with both the cavity and the PDR of NGC 7023, C$_{60}$ has only been observed in the cavity carved by the illuminating star \citep{Berne:2013}. In this paper, we analyze the emission from PAHs and C$_{60}$ using the mid-IR spectra obtained with the IR spectrograph \citep[IRS,][]{Houck:04} in the Short-Low (SL; spectral resolution $\sim$60–128, pixel scale $\sim$1.8$''$; AORs:3871488 and 3871744) and Short-High (SH; spectral resolution $\sim$600, pixel scale $\sim$2.3$''$; AOR:3871232) modules, onboard the Spitzer Space Telescope \citep{Werner:04a}. We cleaned the data using cubism's automatic bad pixel generation with $\sigma_{\rm{TRIM}=7}$ and Minbad-fraction = 0.5 and 0.75 for the global and record bad pixels respectively. We did not apply background subtraction. We extracted spectra from five positions within the nebula at increasing distances from the illuminating star (see Fig.~\ref{fig:spatial_map_NGC_7023}). The five positions are at projected distances of 13.2$''$, 20.1$''$, 27.4$''$, 34.7$''$, and 42.2$''$ from the star with position 1 being the closest and position 5 being the farthest away from the star.

\section{Application of the charge distribution based emission model to PAHs and fullerenes}
\label{sec:Method}

In Paper I, we presented an emission model of PAHs that first calculates the charge distribution of a given PAH molecule and then uses it to calculate the emission spectrum in an astrophysical environment by weighing the spectrum of each charge state of a PAH molecule with its charge fraction and then summing them to obtain the total emission spectrum. By applying our emission model to five different PDRs with varying physical conditions, we demonstrated that the charge distribution emission model could successfully explain the characteristics of the PAH fluxes. In this paper, we extend the application of the charge distribution-based emission model to calculate the emission spectrum of C$_{60}$. In this section, we first describe the calculation and the results of the charge distribution of C$_{60}$ over a range of physical conditions characterized by the ionization parameter $\gamma$, followed by a discussion of the emission spectra of C$_{60}$ in various charge states. We refer the reader to section 3 in Paper I for the analogous discussion on PAHs. 

\subsection{Charge distribution of C$_{60}$}
\label{subsec:charge_distribution_C60}
We calculated the charge distribution using the model described in Paper I. Essentially, the model calculates the charge distribution of molecular species using the principle of ionization balance between the photo-ionization rate and the electron recombination rate (or the electron attachment rate for anions), eventually yielding the following equation.

\begin{equation}
    f(Z) = \frac{k_{\rm e}(Z+1)}{k_{\rm ion}(Z)}f(Z+1)
    \label{eq:frac_PAHs}
\end{equation}

\noindent where  $f(Z)$ is the fraction of molecular species in a charge state $Z$, $k_{\rm ion}(Z)$ and $k_{\rm e}(Z)$ are the photo-ionization and electron recombination rates of a molecule in a charge state $Z$ in units of s$^{-1}$. For $Z = -1$, $k_{\rm e}(Z+1)$ is the electron attachment rate, $k_{\rm ea}(Z)$. Applying the normalization condition, i.e. the sum of the fraction of all the charge states is equal to 1, yields the charge distribution of a molecule. For a detailed description of the model, see Paper I.

The charge distribution calculation relies on the various rates ($k_{\rm ion}(Z)$, $k_{\rm e}(Z)$, and $k_{\rm ea}(Z)$), which depend on the molecular properties of chemical species as well as the physical conditions of the astrophysical environment. The specific rates adopted for PAHs are described in Paper I. Here we briefly describe our calculations of the rates for C$_{60}$.

\subsubsection{Photo-ionization rate}
\label{subsec:photo_ionization_rate}
The photo-ionization rate ($k_{\rm ion}(Z)$) for C$_{60}$ is calculated using the same expression as for PAHs using the molecular properties of C$_{60}$ instead of PAHs. Calculating the ionization rate requires the following molecular parameters: the ionization potential (IP), the absorption cross-section, and the ionization yield.

\textbf{The Ionization Potential}: Table~\ref{tab:Ionization_potential} lists the IPs of the C$_{60}$ molecule adopted in this work. The IP dictates the number of charge states accessible to a molecule. Since we are applying the model to the cavity in NGC~7023 where hydrogen (H) ionizing photons (i.e. the photons with energy > 13.6 eV) are absent, the charge state $Z+1$ will be accessible to C$_{60}$ if the IP($Z$) is less than 13.6 eV. Based on the IPs listed in Table~\ref{tab:Ionization_potential}, the highest possible charge state accessible to C$_{60}$ in the cavity of NGC~7023 is $Z$ = 2. 

\begin{table}
\caption{Ionization potential of C$_{60}$ in various charge states.}
    \centering
    \begin{threeparttable}[t]
    \begin{tabular}{c c}
    \hline
    Charge state ($Z$) & IP($Z$) [eV] \\
    \hline
    -1 & 2.65 \tnote{1,2} \\
     0 & 7.6 \tnote{3}\\
     1 & 11.5 \tnote{4}\\
    \hline
    \end{tabular}
    \begin{tablenotes}[para]
     \item[1] \citet{Stochkel:2013}
     \item[2] \citet{Huang:2014}
     \item[3] \citet{Yoo:1991}
     \item[4] \citet{Dresselhaus:1996}
   \end{tablenotes}
   \end{threeparttable}

\label{tab:Ionization_potential}
\end{table}

\textbf{Absorption cross-section}: Several experimental measurements of the photo-absorption cross-section exist in the literature \citep{Berkowitz:1999, Yasumatsu:1996}. These measurements, however, are inconsistent with one another. \citet{Kafle:2008} identified and attributed the discrepancies to the use of inconsistent vapour pressure data of C$_{60}$ in the analysis of these measurements. \citet{Kafle:2008} compiled the photo-absorption cross-section data from 3.5-26 eV accounting for the necessary changes in the vapour pressure data, which we adopt in this work (see Fig.~\ref{fig:cross-section}).

\textbf{The ionization yield}: The ionization yield for C$_{60}$ has been calculated analytically by \citet{Yasumatsu:1996} using their measurements of the photo-absorption cross-section. However, we did not use their ionization yield data because of the discrepancies in their photo-absorption cross-section, as stated above. Instead, we adopted the semi-empirical relation derived by \citet{Jochims:1996} for calculating the ionization yield of PAHs to calculate the yield for C$_{60}$ (see Fig.~\ref{fig:yield}), assuming the ionization behaviour of C$_{60}$ and PAHs is similar \citep{Tielens:2021}.

\subsubsection{Electron recombination rate}
\label{subsec:electron_recombination_rate}
Since there is no experimentally measured value for the electron recombination rate coefficients for C$_{60}$, we estimated it using the following expression (see equation 8.106 in \citet{Tielens:2021}) which is based on the collisional interaction between an electron and spherical species of radius $r$ = 0.9 N$_{\rm C}^{1/2}$:

\begin{equation}
    k_{\rm e}(Z) = 1.3 \times 10^{-6} Z \,\, \text{N}_{\rm C}^{1/2}\left (\frac{300}{\rm T_{\rm gas}}\right )^{1/2}\rm n_{\rm e}\;\;\;\;\; (\text{s}^{-1})
    \label{eq:recombination_rate}
\end{equation}

\noindent where N$_{\rm C}$ = 60 are the number of the carbon atoms in a molecule, $\rm T_{\rm gas}$ is the gas temperature in K, and $\rm n_{\rm e}$ is the density of electrons in an astrophysical environment in units of cm$^{-3}$. We note that our adopted value of the electron recombination rate matches well with the recent theoretical estimates by Zettergren et al. (in preparation) using the models in \citet{Zettergren:2012} and \citet{Fredrik:2016}.

\subsubsection{Electron attachment rate}
\label{subsec:electron_attachment_rate}
The electron attachment rate coefficient has been determined experimentally as $10^{-6}$ cm$^{3}$ s$^{-1}$ by \citet{Viggiano:2010}. We adopt this experimentally measured value and use the following expression to estimate the electron attachment rate in a given environment:   

\begin{equation}
    k_{\rm ea} = 10^{-6} \rm n_{\rm e} \;\;\;\;\;(\text{s}^{-1})
\end{equation}

\begin{figure}
    \centering
    \includegraphics[scale=0.45]{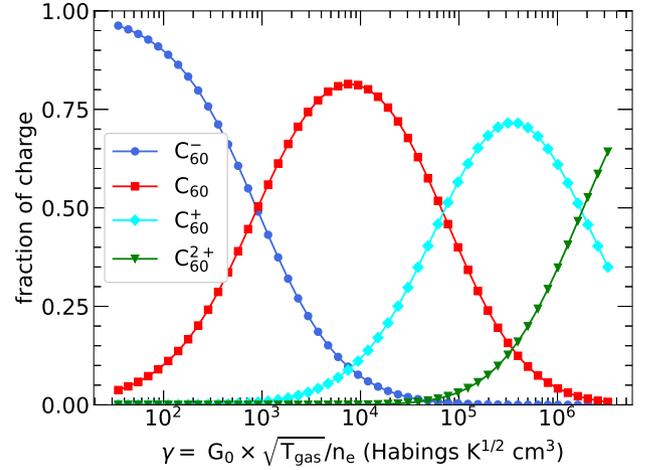}
    \caption{Charge distribution of C$_{60}$ as a function of the ionization parameter, $\gamma$. See text for the details of the calculation.}
    \label{fig:charge_C60}
\end{figure}

\bigskip
Fig.~\ref{fig:charge_C60} shows the charge distribution of C$_{60}$ as a function of the ionization parameter, $\gamma$ = $\rm G_{0} \times \sqrt{\rm T_{\rm gas}} /\rm n_{\rm e}$ Habings K$^{1/2}$ cm$^{3}$. We point out that since the units of $\gamma$ are long, we omit them for the remainder of our paper. The dense NW PDR in NGC~7023 is characterized by a radiation field of G$_{0}$ = 2600 in the units of the Habings field \citep{Chokshi:88}, T$_{\rm gas}$ = 400 K \citep{Fuente:1999, Fleming:2010}, an electron abundance of 1.6 $\times 10^{-4}$ \citep{Sofia:2004} relative to the density of hydrogen, and  gas density, n$_{\rm gas}$, of $10^{4}$ cm$^{-3}$ \citep{Chokshi:88, Joblin:10, Kohler:2014, Beranrd:2015, Joblin:18}, corresponding to a $\gamma = 3.3 \times 10^{4}$ Habings K$^{1/2}$ cm$^{3}$. We expect the cavity region to have a higher FUV field, therefore higher G$_{0}$, but to be much less dense. Hence, we have evaluated the charge distribution for $\gamma$ ranging from 30 to $5 \times 10^{6}$ Habings K$^{1/2}$ cm$^{3}$. We observe that anions dominate in low $\gamma$ regions ($\gamma <$ $10^{3}$), neutrals dominate in intermediate $\gamma$ regions ($10^{3}<\gamma<7\times 10^{4}$) and cations dominate in high $\gamma$ regions ($\gamma >7\times10^{4}$). It is worth pointing out that the general trend of the charge distribution of C$_{60}$ is similar to that of PAHs (Figs. 2 and C1 in Paper I), although the precise numbers of the fraction of molecules in a given charge state vary.

\subsection{Intrinsic Spectra of C$_{60}$}
\label{subsec:intrinsic_spectra_C60}
\begin{figure}
    \centering
    \includegraphics[scale=0.35]{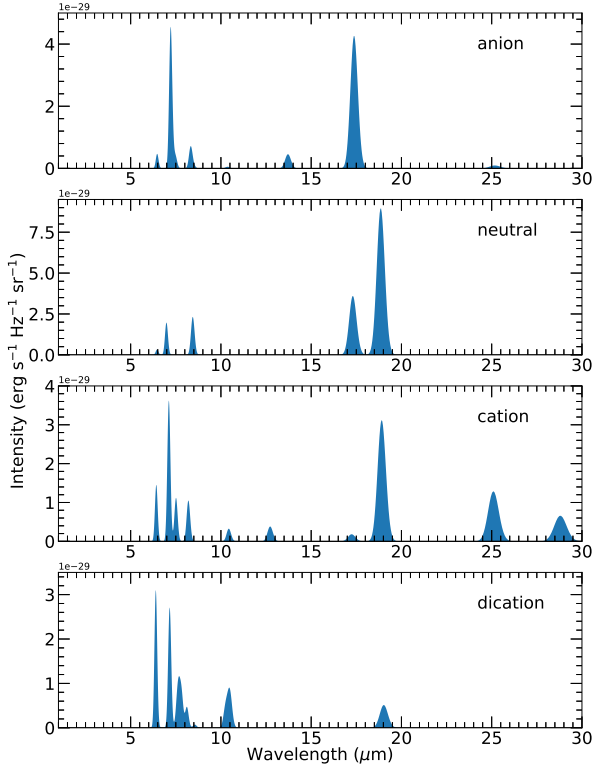} 
    \caption{Emission spectra of C$_{60}$ in the anionic, neutral, cationic, and dicationic states for the excitation conditions of the NW PDR in NGC~7023.}
    \label{fig:intrinsic_spectra_fullerenes}
\end{figure}

\begin{table}
\caption{Average energy (E$_{\rm avg}$) absorbed by each charge state of C$_{60}$ molecule in NGC~7023.}
    \centering
 
    \begin{tabular}{c c}
    \hline
    Charge state ($Z$) & E$_{\rm avg}$($Z$) [eV] \\
    \hline
    -1 & 5.65 \\
     0 & 6.22 \\
     1 & 6.47 \\
     2 & 6.50 \\
    \hline
    \end{tabular}

\label{tab:average_energy}
\end{table}

We calculated the emission spectra of various charge states of C$_{60}$ in NGC~7023 using the emission model presented in Paper I. We begin by calculating the average photon energy absorbed by the C$_{60}$ molecule in different charge states (see equation 12 in Paper I). The resulting average energies are presented in Table~\ref{tab:average_energy}. Following absorption of a photon, a molecule dissipates the absorbed energy by redistributing it across its various vibrational modes, eventually cascading down to the ground state by emitting in the IR. In our model, we include the possibility of the absorption of another photon while the molecule fully cascades down to the ground state. For a thorough description of the emission model, we refer the reader to section 2.2 of Paper I. Our model requires the photo-absorption cross-section and the frequencies and corresponding intensities of the vibrational modes of C$_{60}$ in a given charge state as input molecular parameters to model the C$_{60}$ emission in various charge states. The measurements of the photo absorption cross-section adopted in this work are described in section~\ref{subsec:photo_ionization_rate}. We obtained the frequencies and the intensities of the vibrational modes of C$_{60}$ in various charge states from \citet{Strelnikov:2015}. These authors measured the frequencies of the vibrational modes and the corresponding intrinsic intensities in a Neon matrix. Previous studies have shown that matrix interactions results in uncertainties in frequencies of 0.21\% $\pm$ 0.63\% \citep{Jacox:2002} and this accuracy is born out for PAHs \citep{Mackie:2018}. Intensities of PAH vibrational transitions measured in matrix isolation studies agree within 18\% $\pm$ 62\% with the results of anharmonic density functional theory calculations \citep{Mackie:2018}. We expect the uncertainties for C$_{60}$ to be of the same order.

In Fig.~\ref{fig:intrinsic_spectra_fullerenes}, we present the calculated emission spectra of C$_{60}$ in the anionic, neutral, cationic, and dicationic charge states for absorbed photon energies listed in Table~\ref{tab:average_energy}. Here, we have convolved the modelled emission intensities with a Gaussian profile having a full width at half maximum, corresponding to a spectral resolution of 200, similar to Paper I. We note that the C$_{60}$ anions, cations, and dications exhibit a rich spectrum with various features in the 5-30 $\mu$m region. The relative intensities of the features in the 5-10 $\mu$m versus 15-20 $\mu$m range for all C$_{60}$ charge states are quite interesting. While neutral C$_{60}$ has weak features in the 5-10 $\mu$m range compared to the 15-20 $\mu$m range, the anionic, cationic, and dicationic C$_{60}$ behave differently. The charged species have stronger features in the 5-10 $\mu$m range compared to the features in the 15-20 $\mu$m range, with the effect being most pronounced for C$_{60}$ dications. This demonstrates that the charge state of C$_{60}$ in the spectra observed in the interstellar medium can be probed by comparing the relative intensity of features in the 5-10 $\mu$m range to those in the 15-20 $\mu$m range. To this end, it is worth pointing out that PAHs also exhibit features in the mid-IR that are typically much stronger than the C$_{60}$ features in interstellar spectra owing to the low abundance of C$_{60}$ in the interstellar medium compared to PAHs \citep{Berne:2012, Omont:2016, Cami:2018, Omont:2021}.

\section{Comparison of the model results with the observations}
\label{sec:comparison}

\begin{figure*}
    \centering
    \begin{tabular}{cc}

    \includegraphics[scale=0.40]{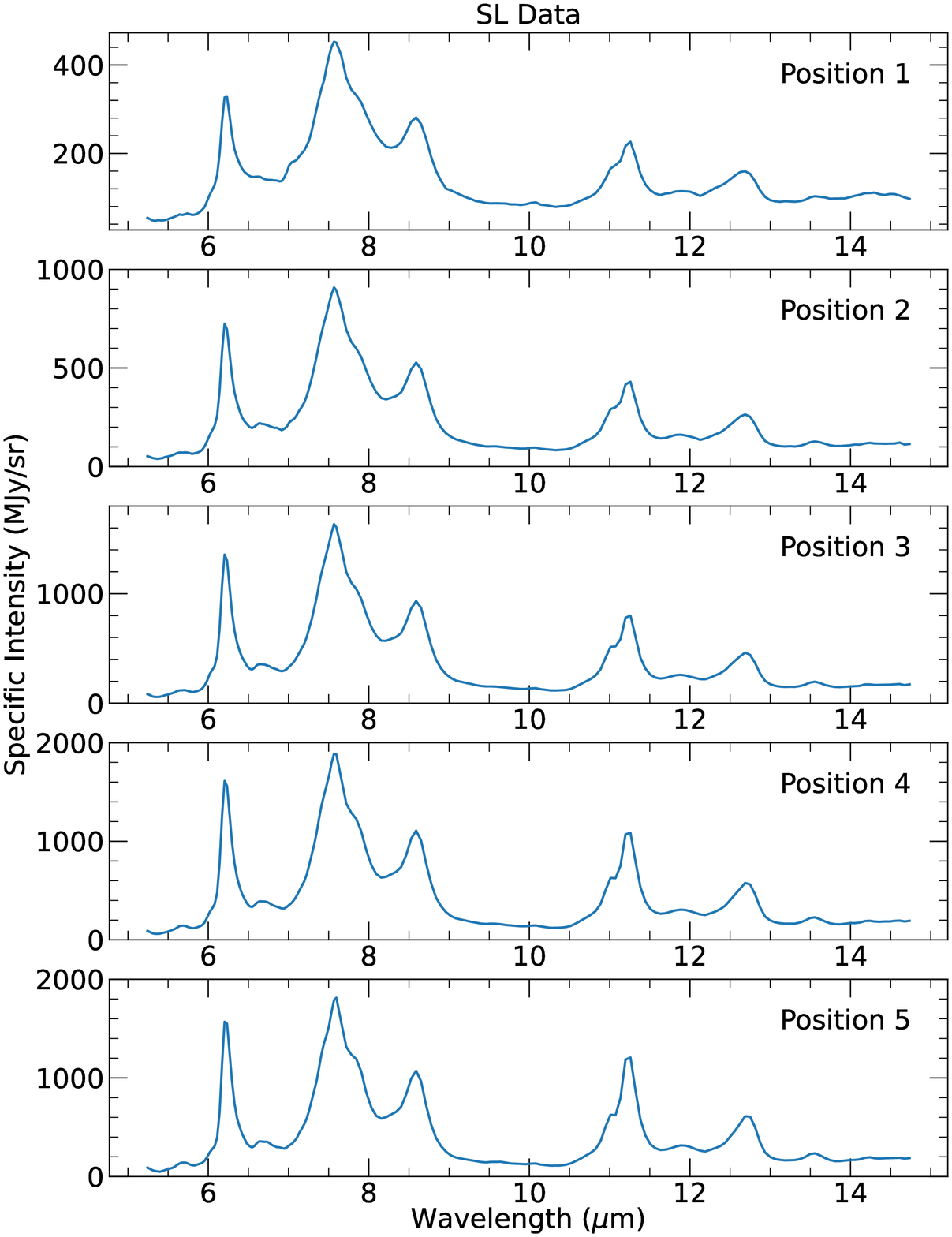} &  \includegraphics[scale=0.40]{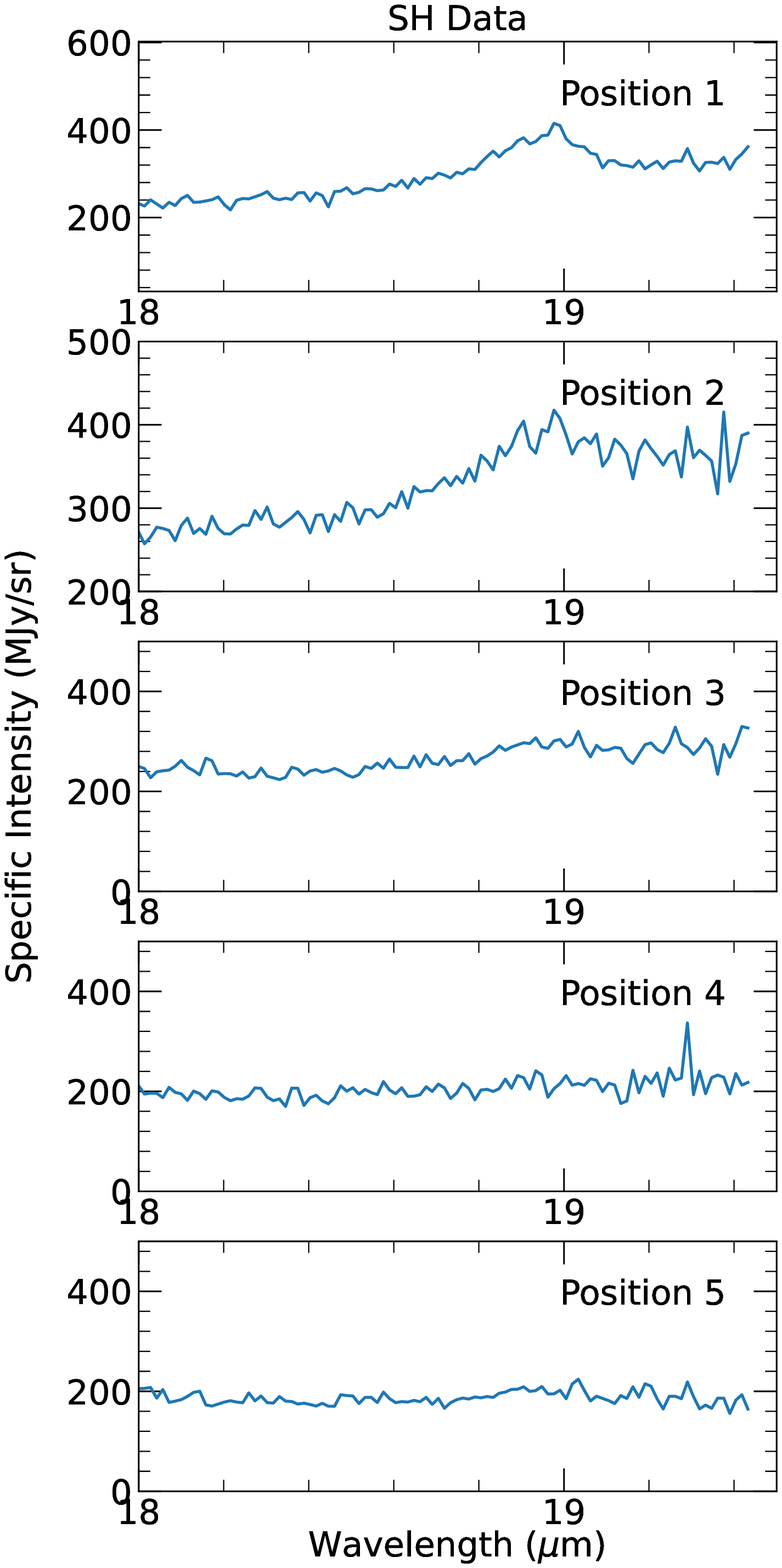}\\
    \end{tabular}
    \caption{The SL and SH spectra observed at five positions in the cavity of NGC~7023. While the PAH bands at 6.2 and 11.2 $\mu$m are observed in all five positions, the C$_{60}$ band at 19.0 $\mu$m is only observed in positions 1 and 2. }
    \label{fig:SL_SH_data}
\end{figure*}  

\begin{figure}
    \centering
    \includegraphics[scale=0.45]{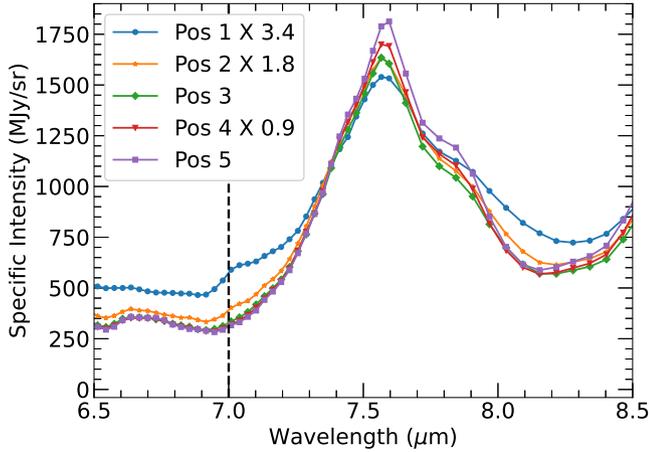}
    \caption{Zoom in of the SL data in the 6.5-8.5 $\mu$m region for all five positions. For clear identification of the 7.0 $\mu$m C$_{60}$ band, we scaled the spectra in positions 1, 3, and 4 matching the blue wing of the 7.7 $\mu$m band roughly in the 7.35 to 7.5 $\mu$m range. For reference, we have indicated the position of the 7.0 $\mu$m C$_{60}$ band.}
    \label{fig:7_micron_band}
\end{figure}

In Paper I, we applied our emission model to a single position in a given PDR. Here, we apply our model to five different positions within one object. In this section, we compare our model results to the PAH and C$_{60}$ observations in the cavity of NGC~7023. In particular, we compare the modelled intensities of the 6.2/11.2 ratio for PAHs and 7.0/19.0 for C$_{60}$ to the observed ratios. First, in section~\ref{subsec:observations}, we describe the observations of PAHs and C$_{60}$ in the cavity of NGC~7023 and the procedure we used to extract fluxes of the 6.2, 7.0, 11.2, and 19.0 $\mu$m bands. We then compare the observed ratios to the model predictions in sections~\ref{subsec:model_results_PAH_observations} and \ref{subsec:model_results_C60} for PAHs and C$_{60}$ respectively.

\subsection{Observations of PAHs and C$_{60}$ in NGC~7023}
\label{subsec:observations}

\begin{table}
\caption{Values of the PAH ratio 6.2/11.2 and C$_{60}$ ratio 7.0/19.0 in the cavity of NGC~7023.}
    \centering
 
    \begin{tabular}{c c c}
    \hline
    Position & PAH ratio 6.2/11.2 & C$_{60}$ ratio 7.0/19.0 \\
    \hline
    1 & 3.92 & 1.04\\
    2 & 3.87 & 0.88\\
    3 & 3.70 & --\\
    4 & 3.23 & --\\
    5 & 2.77 & -- \\
    \hline
    \end{tabular}

\label{tab:observed_values_ratios}
\end{table}

We considered five positions in the cavity of NGC~7023 at increasing distances from the illuminating star (see Fig.~\ref{fig:spatial_map_NGC_7023}). In Figs.~\ref{fig:SL_SH_data} and ~\ref{fig:7_micron_band}, we show the mid-IR spectra observed in the SL and the SH modules for the five positions. While the 6.2 and 11.2 $\mu$m PAH bands are observed in all five positions, the 7.0 and 19.0 $\mu$m C$_{60}$ bands are only visible in positions 1 and 2. The 19.0 $\mu$m C$_{60}$ band is easily identified in positions 1 and 2 (see the right panel of Fig.~\ref{fig:SL_SH_data}); however, the 7.0 $\mu$m band is blended with the 7.7 $\mu$m PAH band and thus its characteristics are challenging to extract. To demonstrate the presence of the 7.0 $\mu$m band in positions 1 and 2, we present a zoom-in of the spectra in the 6.5-8.5 $\mu$m region for the five positions in Fig.~\ref{fig:7_micron_band}. Furthermore, to distinguish the C$_{60}$ band from the wing of the strong 7.7 $\mu$m PAH band, we match the blue wing of the latter in the 7.35 - 7.5 $\mu$m region by scaling the spectra of position 1, 2, and 4 by a factor of 3.4, 1.8, and 0.9 respectively. After scaling, we find that the spectra of positions 3, 4, and 5 match in the 6.5-7.5 $\mu$m range, but not for positions 1 and 2, where we can now clearly observe the 7.0 $\mu$m band of considerable and weak strength in positions 1 and 2 respectively.

To obtain the fluxes of the 6.2 and 11.2 $\mu$m PAH bands, we first subtracted a spline continuum with anchor points roughly at 5.4, 5.8, 5.9, 6.9, 8.1, 8.9, 9.0, 9.1, 9.3, 9.5, 9.8, 10.2, 10.4, 11.6, 12.2, 13.2, 13.8 $\mu$m. Then we integrated the continuum subtracted spectra in the 5.9-6.9 and 10.4-11.6 $\mu$m regions to extract fluxes for the 6.2 and 11.2 $\mu$m bands respectively. Table~\ref{tab:observed_values_ratios} shows the resulting PAH ratio of 6.2/11.2.

\begin{figure*}
    \centering
    \begin{tabular}{cc}

    \includegraphics[scale=0.40]{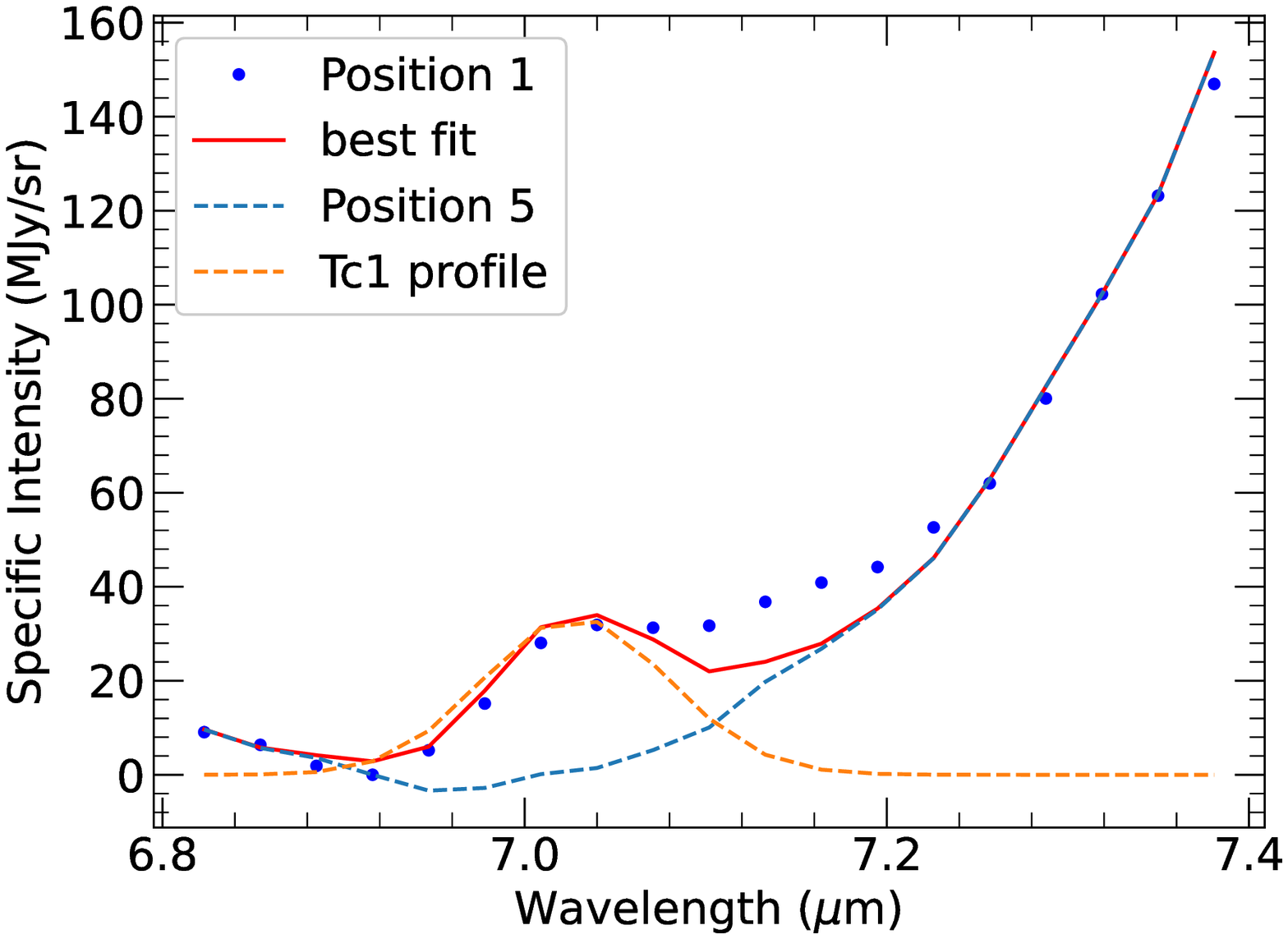} &  \includegraphics[scale=0.40]{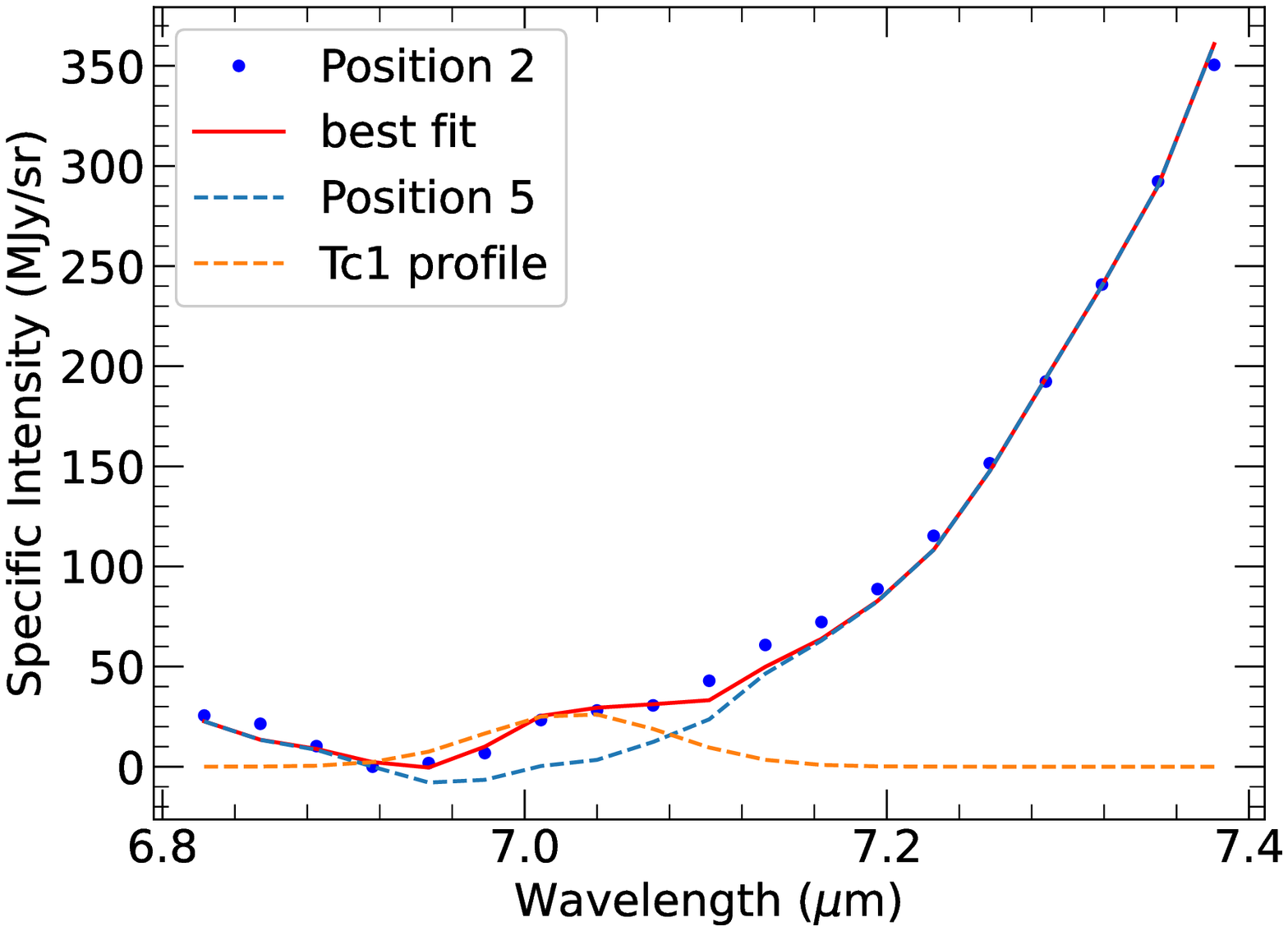}\\
      \includegraphics[scale=0.40]{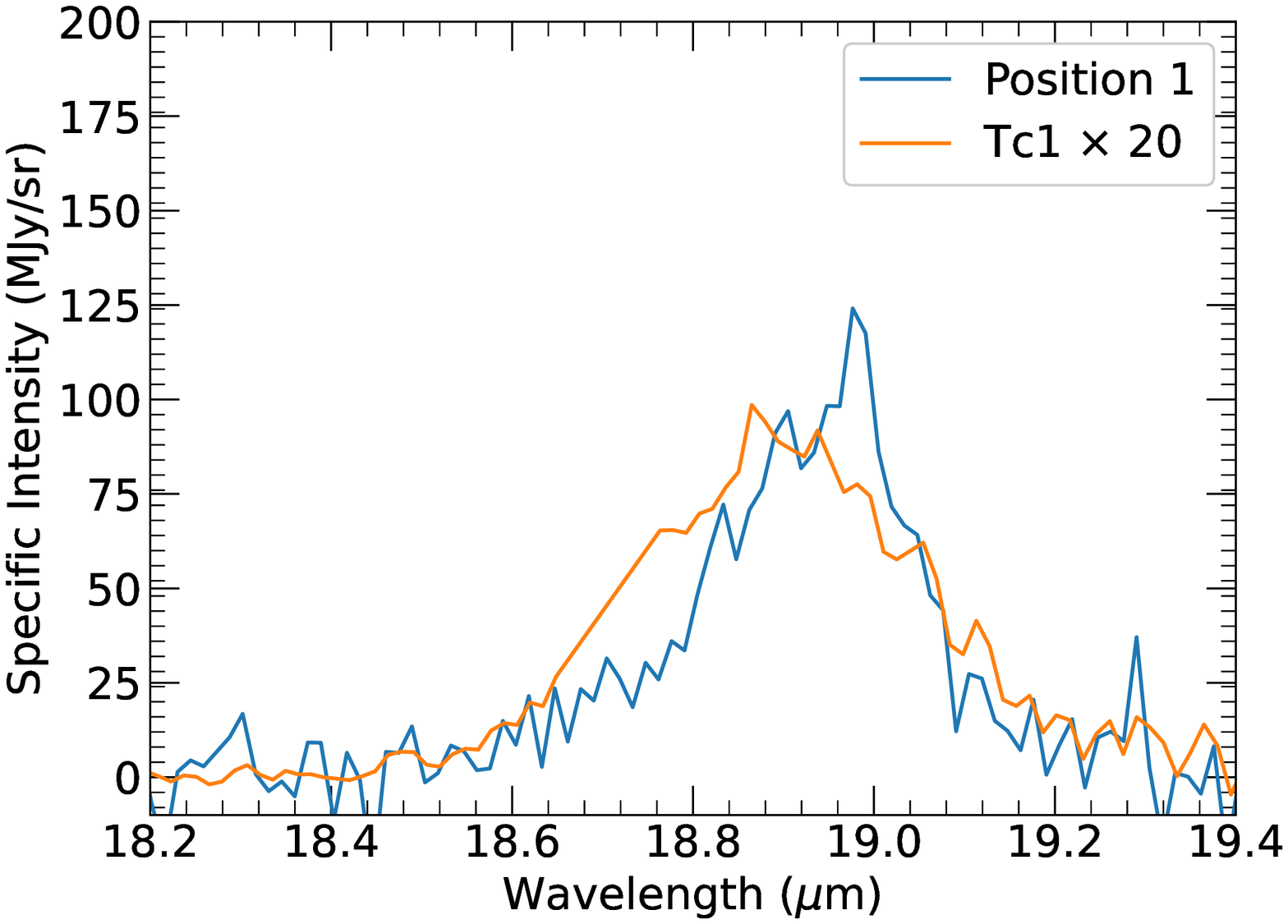} &  \includegraphics[scale=0.40]{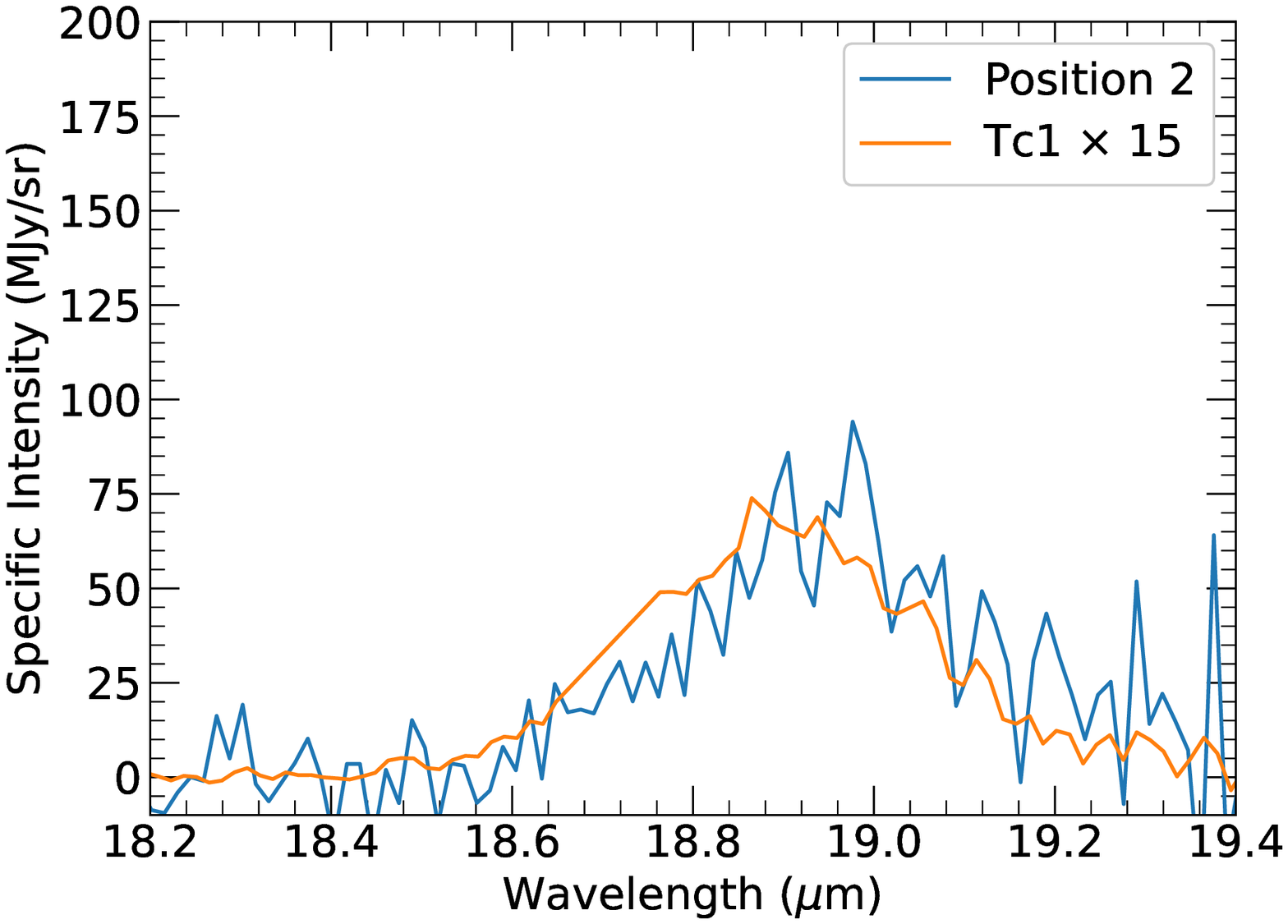}\\
    \end{tabular}
    \caption{Top panel: The fit in the 6.8-7.4 $\mu$m region to extract the 7.0 $\mu$m C$_{60}$ band in positions 1 and 2. The spectral region is fit with the 7.0 $\mu$m C$_{60}$ band as observed in Tc1 and the blue wing of the 7.7 $\mu$m PAH band from position 5. Bottom panel: The 19.0 $\mu$m C$_{60}$ in positions 1 and 2 along with the scaled 19.0 $\mu$m C$_{60}$ band as observed in Tc1. }
    \label{fig:SL_SH_data_fits}
\end{figure*}

The flux of the 7.0 $\mu$m C$_{60}$ band is challenging to extract because it blends with the 7.7 $\mu$m PAH band, which is lying on a rising continuum. Therefore, we first isolate the 7.0 $\mu$m C$_{60}$ band by subtracting the contribution from the PAHs and the continuum to the spectrum. To accomplish this, we fit the 6.8-7.4 $\mu$m region in the continuum subtracted spectra with the profile of the 7.0 $\mu$m C$_{60}$ band as observed in the planetary nebula Tc1 by \citet{Cami:2010} and the blue wing of the 7.7 $\mu$m band from position 5 representing the contribution of PAHs (see top panels in Fig.~\ref{fig:SL_SH_data_fits}). We note that in position 5, we do not observe any C$_{60}$ band; thus, we interpret the 6.8-7.4 $\mu$m region spectrum in position 5 as representing the contribution from PAHs. Finally, we integrate the fitted 7.0 $\mu$m Tc1 profile to obtain the flux of the 7.0 $\mu$m C$_{60}$ band in positions 1 and 2. For the flux of the 19.0 $\mu$m C$_{60}$ band, we subtract a spline continuum with anchor points roughly at 15.1, 15.5, 15.9, 17.7, 18.5, and 19.4 $\mu$m in the SH data. Then, we integrate the continuum subtracted spectra in the 18.5-19.4 $\mu$m region to extract the flux of the 19.0 $\mu$m C$_{60}$ band. We also fitted the profile of the 19.0 $\mu$m band as observed in Tc1, where PAHs are absent, to the SH data observed in NGC~7023. Fig.~\ref{fig:SL_SH_data_fits} shows the continuum subtracted 19.0 $\mu$m band in NGC~7023 and the scaled 19.0 $\mu$m band in Tc1. Table~\ref{tab:observed_values_ratios} lists the C$_{60}$ ratio of 7.0/19.0 in positions 1 and 2.

Perusing the spectra in Fig.~\ref{fig:SL_SH_data} and the intensity ratios in Table~\ref{tab:observed_values_ratios}, we recognize that the variations in PAH emission features are very modest but systematic: as the positions approach the star, the 6.2/11.2 PAH ratio increases from 2.77 to 3.92. The 7.0/19.0 C$_{60}$ ratio also increases from position 2 to 1 (Table~\ref{tab:observed_values_ratios}). As we will discuss later, this suggests that the $\gamma$ value in the cavity does not vary drastically.

\subsection{Model results versus PAH observations}
\label{subsec:model_results_PAH_observations}
\begin{figure}
    \centering
    \includegraphics[scale=0.45]{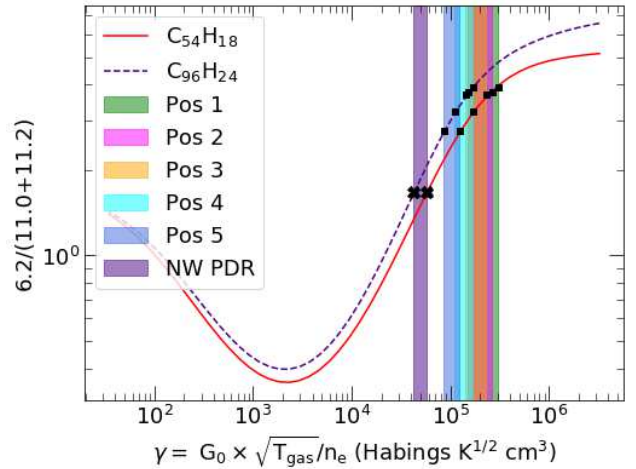}
    \caption{The modelled 6.2/11.2 PAH ratio for circumcoronene (solid red curve) and circumcircumcoronene (dashed blue curve) with observed ratios overplotted. The square markers correspond to the ratio observed at five positions in the cavity, and the cross marker corresponds to the ratio observed at the NW PDR. The shaded colored region represents the range in $\gamma$ values obtained from the comparison between observations and two theoretical curves.} 
    \label{fig:ratio_plot_PAHs}
\end{figure}

Following Paper I, we modelled the emission of circumcoronene (C$_{54}$H$_{18}$) and circumcircumcoronene (C$_{96}$H$_{24}$) in NGC~7023 and created curves of the 6.2/11.2 PAH ratio as a function of $\gamma$ values. We chose these two compact PAH molecules to represent PAHs in the cavity of NGC~7023 because of their stability that can withstand harsh conditions closer to the star \citep{Andrews:2015, Croiset:2016}. Moreover, since PAHs undergo photoprocessing with changing physical conditions \citep[e.g.][]{Peeters:2017, Murga:2022}, we chose two molecules to represent a range in the size of the molecules consistent with the determined PAH sizes for NGC~7023 derived by \citet{Croiset:2016}. Fig.~\ref{fig:ratio_plot_PAHs} shows the modelled 6.2/11.2 vs $\gamma$ curves with the observed values for the five positions studied in this paper. For reference, we also overplot the observed value of the 6.2/11.2 for the NW PDR (see Appendix~\ref{sec:spec_NW_PDR}). 

For each position, the comparison of the observed 6.2/11.2 PAH ratio with the theoretical curve results in two gamma values, one from circumcoronene and the other from circumcircumcoronene. The $\gamma$ value derived from circumcoronene represents the upper limit, while that derived from circumcircumcoronene represents the lower limit. We find that, in the cavity, the observed values lie in the $\gamma$ ranging from $8.5 \times 10^{4}$ to 3 $\times 10^{5}$ with position 1 exhibiting the highest $\gamma$ value and position 5 exhibiting the lowest $\gamma$ value.  
While the $\gamma$ values predicted from the PAH observations systematically decrease as we go from position 1 to position 5, the change in the $\gamma$ values is small given that the overall range of $\gamma$ values is much wider. This result is consistent with the fact that the spectra of positions 1-5 in the cavity show modest variations (see Fig.~\ref{fig:SL_SH_data}).

We further note that the $\gamma$ values in the five positions of the cavity do not differ significantly from the $\gamma$ value of the dense NW PDR. The predicted $\gamma$ values in the cavity are at most a factor of 7 larger than in the NW PDR, suggesting that the physical conditions in the NW PDR and the cavity do not differ significantly.   

\subsubsection{Physical conditions in NGC~7023}
\label{subsubsec:cavity_geometry}

The comparison of the modelled 6.2/11.2 PAH ratio for circumcoronene with the observed ratio has shown that i) the $\gamma$ values in the cavity only vary by a factor of 3 from position 5 to position 1 and ii) the $\gamma$ values in the cavity and the dense NW PDR do not differ significantly from each other. However, these predicted values of $\gamma$ in the cavity are not consistent with those estimated by \citet{Berne:2015}. These authors estimated $\gamma$ values ranging from 4$\times 10^{4}$ to $\sim$ 7$\times 10^{6}$. \citet{Berne:2015} adopted the projected distance as the location of the emitting PAHs, therefore upon approaching the star, the effect of geometric dilution diminishes (with $r^{-2}$) and the radiation field, G$_{0}$, increases from $\sim$ 7$\times 10^{3}$ Habings at 25$''$ from the star to $\sim$ 2$\times 10^{5}$ Habings at 5$''$ from the star. At the same time, the gas density, n$_{\rm gas}$, was assumed to drop from $\sim$ 10$^{4}$ cm$^{-3}$ to $\sim$ 10$^{3}$ cm$^{-3}$, and the higher abundance of C$^{+}$ of 3$\times 10^{-4}$ was assumed to calculate the electron abundance in the cavity. As a result, in their model, $\gamma$ increases by a factor of $\sim$ 170 from $\sim$ 4$\times 10^{4}$ to $\sim$ 7$\times 10^{6}$ in the cavity.

To explain the significantly lower values of $\gamma$ in the cavity derived in this work, we suggest that the emission from PAHs at the five positions between the dense PDR and the star does not originate from the projected distances. Therefore, the systematic increase in the radiation field, G$_{0}$, caused by geometric dilution as positions approach the star and increase $\gamma$ values, will not be relevant. Similarly, in this scenario, the gas density does not need to systematically decreases from position 1 to 5; the other assumption that resulted in higher $\gamma$ values.

Finally, we note that the observations of C$_{60}$ in positions 1 and 2 can reveal additional information about the change in the physical conditions in the cavity of NGC~7023. \citet{Andrews:2016} modelled the dehydrogenation of circumcoronene and found that the PAH to carbon-cluster transition scales with $G_{0}^{2.5}/n_{\rm e}$. Assuming that C$_{60}$ forms via dehydrogenation of PAHs, we can surmise that the change in the $\gamma$ values from position 5 where only PAHs are observed, to position 1 where both PAHs and C$_{60}$ are observed, is driven more by the changes in G$_{0}$ than by changes in gas density.

\subsection{Model results versus C$_{60}$ observations}
\label{subsec:model_results_C60}
\begin{figure}
    \centering
    \includegraphics[scale=0.45]{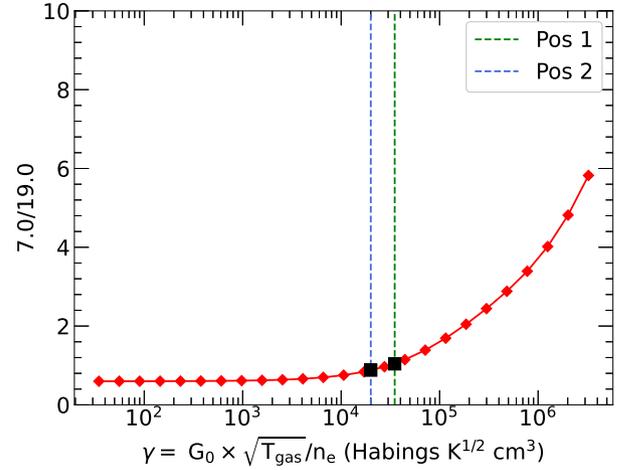}
    \caption{The modelled 7.0/19.0 C$_{60}$ ratio with observed values of the ratio overplotted. The square markers correspond to the ratio observed at positions 1 and 2 in the cavity.}
    \label{fig:ratio_plot_C60}
\end{figure}
We next compare the C$_{60}$ observations with the model calculations to investigate the charge state of C$_{60}$ as well as to check if the $\gamma$ values predicted from C$_{60}$ observations are consistent with those predicted from PAH observations. Similar to PAHs, we create a curve of 7.0/19.0 C$_{60}$ band ratio, and compare the observed values for two positions (see Fig.~\ref{fig:ratio_plot_C60}). We note that the $\gamma$ values for positions 1 and 2 based on C$_{60}$ ratios (3.5 $\times 10^{4}$ for positions 1 and 2 $\times 10^{4}$ for position 2) differ from those based on PAH observations by an order of magnitude. We discuss this disagreement between the $\gamma$ values predicted from C$_{60}$ and PAH ratios further in section~\ref{sec:Discussion}. Nonetheless, the C$_{60}$ predicted $\gamma$ values are also much lower than the previous estimates \citep{Berne:2015} and similar to the NW PDR estimate, further confirming our conclusions that the emission in the cavity is not originating from projected distances (section~\ref{subsubsec:cavity_geometry}).

For the $\gamma$ values obtained in positions 1 and 2 based on C$_{60}$ ratios, we use the Fig.~\ref{fig:charge_C60} obtained from the charge distribution model to determine the charge fraction of C$_{60}$ molecules in positions 1 and 2. We find that for positions 1 and 2, C$_{60}$ is primarily neutral with 68\% neutrals, 29\% cations, and 2\% anions in position 1 and 80 \% neutrals, 14\% cations, and 5\% anions in position 2. In both positions, the dictations have negligible fractions. \citet{Berne:2013} estimated an ionization fraction of 38\% for C$_{60}$ at a projected distance of 7.5$''$ from the star in NGC 7023. In this work, positions 1 and 2 correspond to projected distances of 13.2$''$ and 20.1$''$ from the star. Thus, the theoretical prediction of the ionization fraction based on the charge distribution model matches well with the observations.



\section{Discussion}
\label{sec:Discussion}

The $\gamma$ values in positions 1 and 2 predicted from C$_{60}$ and PAH ratios differ by a factor of 10 (section~\ref{subsec:model_results_PAH_observations} and \ref{subsec:model_results_C60}). In this section, we will discuss two potential explanations for this discrepancy.

The first explanation is based on the uncertainties in the modelled 6.2/11.2 and 7.0/19.0 band ratios, which in turn are a result of the uncertainties in the charge distribution calculations. The charge distribution calculations depend on the ionization and electron recombination/attachment rates of PAHs and C$_{60}$. While the molecular properties involved in the calculations of the ionization rate have been experimentally measured for both PAHs and C$_{60}$, there is no such measurement for the recombination rate of astrophysically relevant PAHs and C$_{60}$. Due to these gaps in laboratory data, we adopt theoretical expressions to calculate recombination/attachment rates. However, these theoretical expressions are only an approximation \citep{Tielens:2008}. These approximations lead to uncertainties in the charge distribution calculations, which further translate into the band ratio calculations, and for this work, result in the systematic uncertainties in the derived $\gamma$ values from the C$_{60}$ and PAH observations. Therefore, similar to Paper I, we highlight that laboratory experiments to determine electron recombination and attachment rates of astrophysically relevant PAHs and C$_{60}$ are highly desirable.

An alternate explanation is that PAHs and C$_{60}$ emission are not co-spatial as the observed emission represents all emission along the line-of-sight, assuming that the uncertainty in the rates for our model calculations is not significant enough to account for the discrepancy in the $\gamma$ values of a factor of 10. This interpretation is supported further by the fact that the C$_{60}$ ionization fractions predicted using the C$_{60}$ derived $\gamma$ values agree well with the observations (see section~\ref{subsec:model_results_C60}), lending confidence in our C$_{60}$ derived $\gamma$ values. We note that the laboratory experiments to determine electron recombination and attachment rates will be useful in eliminating the uncertainty in the rates and validating the argument about the non-cospatial nature of PAH and C$_{60}$ emission. In general, accurate molecular physics parameters carry the potential to transform PAHs and C$_{60}$ observations into astronomers tools to study the Universe.

Finally, we note that, in principle, dehydrogenation can also affect the 6.2/11.2 ratio. However, \citet{Hony:2001} concluded, based on the detailed analysis of the observed bands in the 11-14 $\mu$m region in a wide variety of astronomical objects, that dehydrogenation has little influence on the bands in this region. This is because the transition from fully hydrogenated to fully dehydrogenated PAH species is very sharp in the G$_{0}$/n$_{\rm H}$ratio \citep{Tielens:2005, Andrews:2016}. \citet{Andrews:2016} demonstrated that for G$_{0} > 10^{4}$ Habings (probable radiation field conditions in the cavity of NGC 7023), circumcoronene would fully dehydrogenate, while circumcircumcoronene would remain fully hydrogenated. It is worth pointing out that once the PAH is (almost) fully dehydrogenated, carbon loss and isomerization to fullerene cages occur \citep{Berne:2012, Berne:2015, Zhen:2014A}. In that case, the interstellar PAH family in the cavity of NGC~7023 will be sharply bounded on the side of small-sized PAHs. At that (size) boundary, dehydrogenation will be important and may affect the 6.2/11.2 ratio. However, even only slightly larger PAHs will be fully hydrogenated, and our calculations well represent the expected 6.2/11.2 ratio. As the structures of the partially dehydrogenated PAHs and their IR characteristics have yet to be studied, we refrain from quantifying this aspect. However, we surmise that the effect will only be minor as - at any point - only a very limited range of PAH sizes will be affected.

\section{Summary}
\label{sec:Summary}
We simulated the emission of polycyclic aromatic hydrocarbons (PAHs) and fullerene (C$_{60}$) molecules in the cavity carved out by the star in the reflection nebula NGC~7023 using the charge distribution-based emission model presented in Paper I. The model first independently calculates the charge distribution of a given molecule and the emission spectrum in the various charge states of the molecule in an astrophysical environment using the experimentally measured or theoretically calculated molecular properties. The total IR emission from a molecule is then computed by weighting the spectrum of each charge state of a molecule with its corresponding charge fraction.

We discussed the results of our model's application to PAHs in Paper I. In this paper, we discuss the results of our model's application to C$_{60}$ and compare the C$_{60}$ results with the PAH results. The analysis of the charge distribution of C$_{60}$ as a function of the ionization parameter $\gamma = \rm G_{0}\times \rm T_{\rm gas}^{1/2}/\rm n_{\rm e}$ demonstrates that the ionization behaviour of C$_{60}$ is similar to that of PAHs, with anions dominating in low $\gamma$ regions ($\gamma <$ $10^{3}$), neutrals dominating in intermediate $\gamma$ regions ($10^{3}<\gamma<7\times 10^{4}$), and cations dominating in high $\gamma$ regions ($\gamma >7\times10^{4}$). Using the emission spectrum of the C$_{60}$ in the anionic, neutral, cationic, and dicationic states, we show that the relative intensity of the features in the 5-10 $\mu$m versus 15-20 $\mu$m range can be used as a tracer of the C$_{60}$ charge state.

From the model calculations, we further computed the 6.2/11.2 band ratio for circumcoronene and circumcircumcoronene and 7.0/19.0 for C$_{60}$ as a function of $\gamma$. We compared the modelled ratios to the observed ratios at five positions in the cavity of NGC~7023 and predicted the $\gamma$ values at these positions. The $\gamma$ values derived from PAH ratios at these five positions do not differ significantly. Furthermore, the PAH derived $\gamma$ values are at most a factor of 7 higher than the $\gamma$ value at the dense north-west (NW) PDR in NGC~7023 and much lower than the previous estimates in the cavity by \citet{Berne:2015}. Moreover, we find that the C$_{60}$ derived $\gamma$ values are also much lower than the previous estimate. Based on these findings, we propose that the PAH and C$_{60}$ emission in the cavity does not originate from locations at the observed projected distances. Furthermore, if the C$_{60}$ in positions 1 and 2 in the cavity form from the PAH dehydrogenation, the change in $\gamma$ values in the cavity is driven more by G$_{0}$ rather than by the gas density.

Finally, we note that our C$_{60}$ derived $\gamma$ values are a factor of 10 lower than the PAH derived $\gamma$ values. We discuss two possible explanations for this discrepancy. 
In the first explanation, we attribute the differences in the two $\gamma$ values to uncertainties in the electron recombination rates of PAHs and C$_{60}$ for which we lack experimental measurements. In the second explanation, we argue that the PAHs and C$_{60}$ are likely not co-spatial, resulting in different $\gamma$ values from each model.

\section*{Acknowledgements}
EP and JC acknowledge support from an NSERC Discovery Grant.

\section*{Data Availability}
The data underlying this article will be shared on reasonable request to the corresponding author.


\bibliographystyle{apj}
\bibliography{PAH_papers_2}

\appendix
\appendixpage
\addappheadtotoc

\begin{appendices}

\section{Absorption cross-section}
\label{sec:absorption}
\begin{figure}
    \centering
    \includegraphics[scale=0.35]{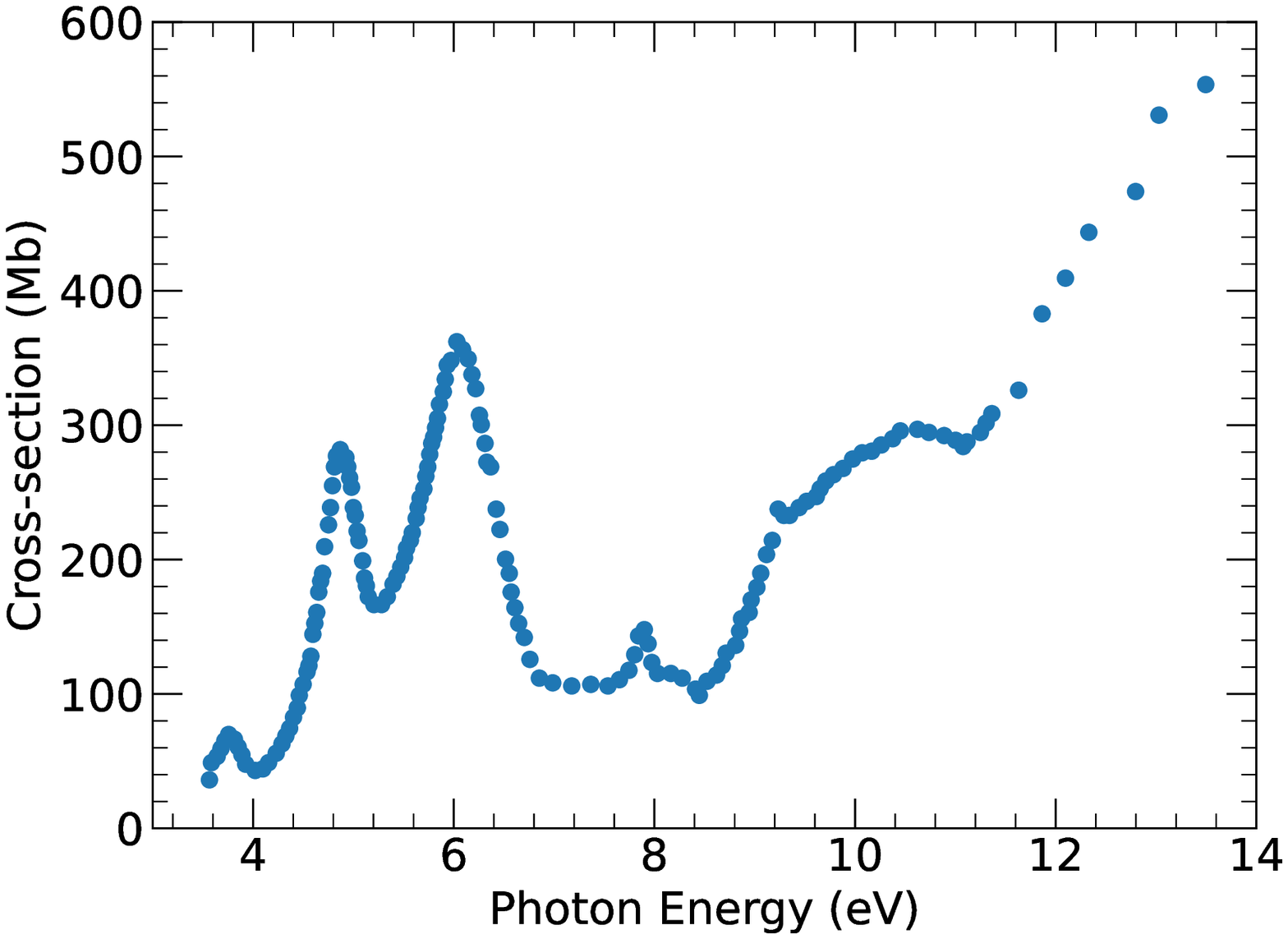}
    \caption{The absorption cross-section of C$_{60}$ from \citet{Kafle:2008} that we adopted in this work.}
    \label{fig:cross-section}
\end{figure}
Fig.~\ref{fig:cross-section} shows the absorption cross-section from \citet{Kafle:2008} that we adopted in this work. We note that we used the same absorption cross-section for all  C$_{60}$ charge states. 

\section{Ionization yield}
\label{sec:yield}
\begin{figure}
    \centering
    \includegraphics[scale=0.35]{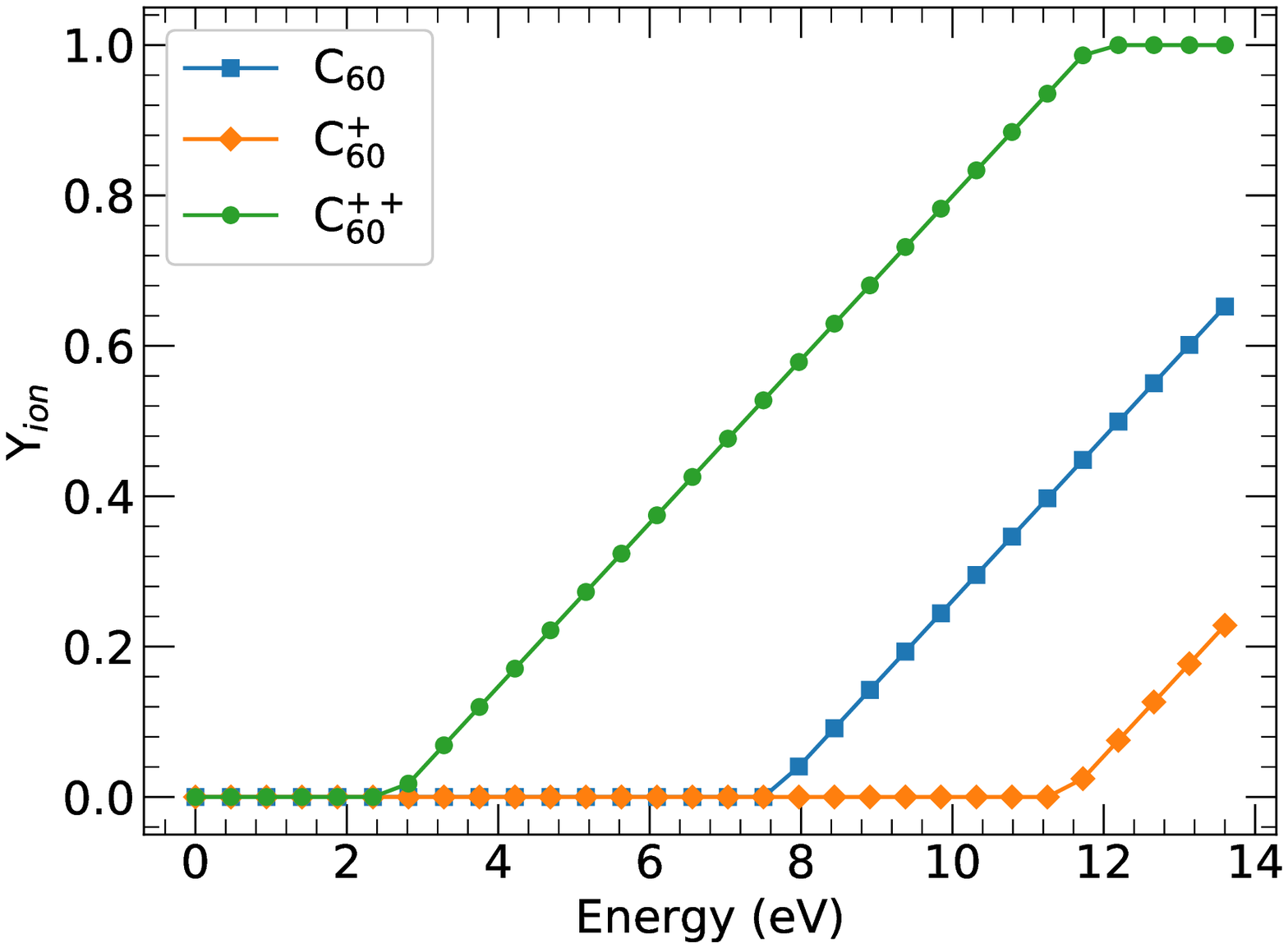}
    \caption{The ionization yield of various charge states of C$_{60}$ adopted in this work.}
    \label{fig:yield}
\end{figure}
Fig.~\ref{fig:yield} shows the ionization yield we used for each charge state of C$_{60}$ derived from the formalism of \citet{Jochims:1996}.

\section{Spectrum at the NW PDR}
\label{sec:spec_NW_PDR}
\begin{figure}
    \centering
    \includegraphics[scale=0.35]{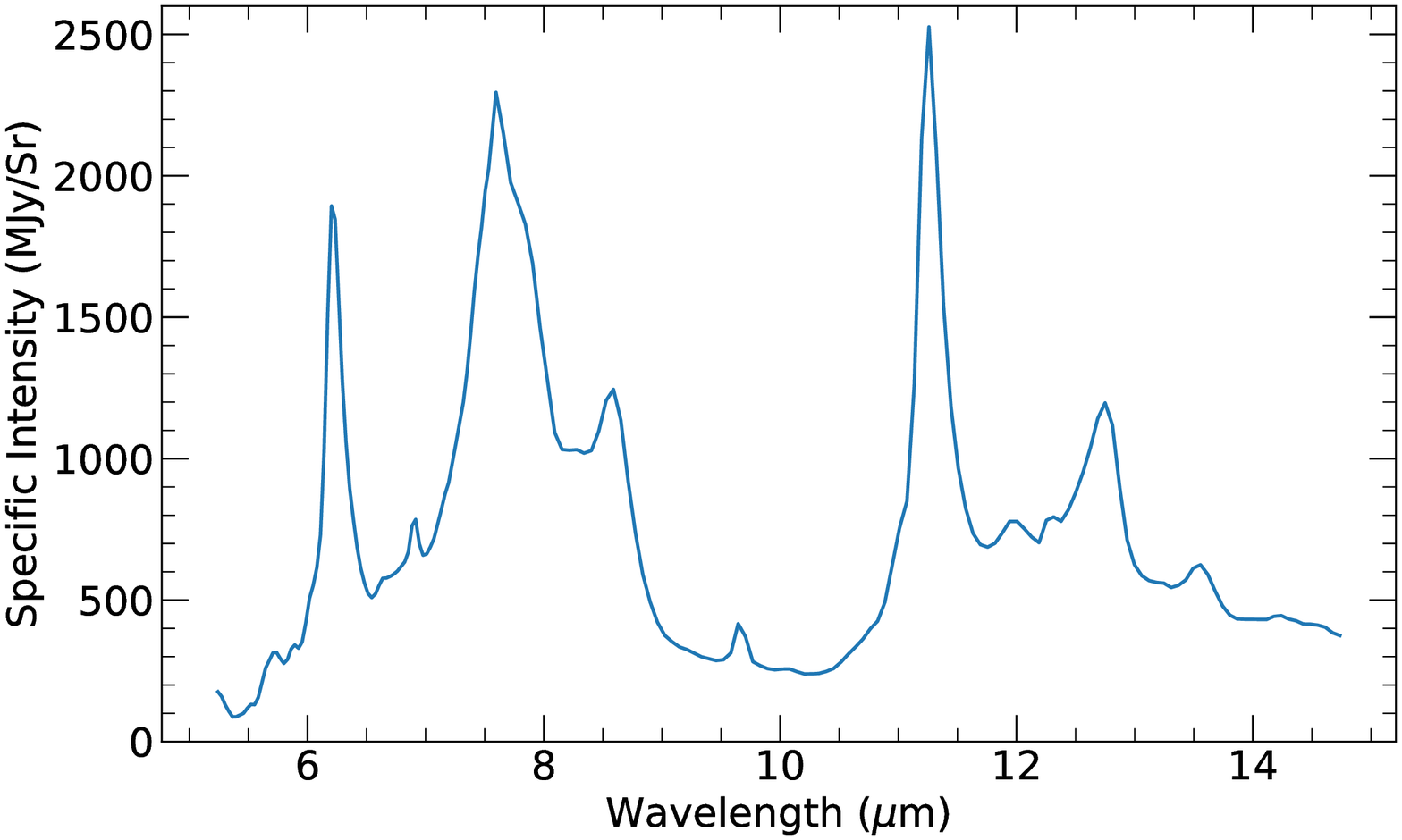}
    \caption{The SL spectrum observed at the NW PDR front in NGC~7023}
    \label{fig:spectra_PDR}
\end{figure}
Fig.~\ref{fig:spectra_PDR} shows the spectrum observed in the SL module at the NW PDR front in reflection nebula NGC~7023. The PAH bands at 6.2 and 11.2 $\mu$m are clearly observed in the spectrum. For this spectrum, the PAH ratio of 6.2/11.2 is 1.68.

\end{appendices}

\bsp	
\label{lastpage}
\end{document}